\documentclass[doublecol]{epl2} 

\usepackage{graphicx}
\usepackage{float}
\usepackage{amsmath, amssymb}
\usepackage{lmodern}
\usepackage[T1]{fontenc}
\usepackage[utf8]{inputenc}
\usepackage[english]{babel}
\usepackage{color}

\newcommand\lsim{\mathrel{\rlap{\lower4pt\hbox{\hskip1pt$\sim$}}
    \raise1pt\hbox{$<$}}}
\newcommand\gsim{\mathrel{\rlap{\lower4pt\hbox{\hskip1pt$\sim$}}
    \raise1pt\hbox{$>$}}}

\title{Role of predator-prey reversal in Rock-Paper-Scissors models}
\shorttitle{...} 

\author{P.P. Avelino\inst{1,2}, B.F. de Oliveira\inst{3}, and R.S. Trintin\inst{3}}
\shortauthor{Avelino, de Oliveira, Trintin}

\institute{\inst{1}Departamento de F\'{\i}sica e Astronomia, Faculdade de Ci\^encias, Universidade do Porto, Rua do Campo Alegre s/n, 4169-007 Porto, Portugal\\
\inst{2}Instituto de Astrof\'{\i}sica e Ci\^encias do Espa{\c c}o, Universidade do Porto, CAUP, Rua das Estrelas, 4150-762 Porto, Portugal\\
\inst{3}Departamento de F\'\i sica, Universidade Estadual de Maring\'a, 87020-900 Maring\'a, PR, Brazil\\
}


\abstract{
In this letter we consider a single parameter generalization of the standard three species Rock-Paper-Scissors (RPS) model allowing for predator-prey reversal. This model, which shall be referred to as $\kappa$RPS model, incorporates bidirectional predator-prey interactions between all the species in addition to the unidirectional predator-prey interactions of the standard RPS model. We study the dynamics of a May-Leonard formulation of the $\kappa$RPS model using lattice based spatial stochastic simulations with random initial conditions.  We find that if the simulation lattices are sufficiently large for the coexistence of all three species to be maintained, the model asymptotically leads to the formation of spiral patterns whose evolution is qualitatively similar to that of the standard RPS model, albeit with larger characteristic length and time scales. We show that there are in general two distinct scaling regimes: one transient curvature dominated regime in which the characteristic length of the population network grows with time and another where it becomes a constant. We also estimate the dependence of the asymptotic value of the characteristic length of the population network on the likelihood of predator-prey reversal and show that if the simulation lattices are not sufficiently large then predator-prey reversal can have a significant negative impact on coexistence. Finally, we interpret these results by considering the much simpler dynamics of circular domains.}

\begin{document}

\maketitle
\section{Introduction}

The modeling of population dynamics in simple biological systems often relies on spatial stochastic predator-prey models where populations of different species are involved in interspecific interactions. The spatial stochastic Rock-Paper-Scissors (RPS) model, perhaps the most famous model of this type, describes the spatial dynamics of a population of three species subject to reproduction, mobility and cyclic predator-prey interactions \cite{1996-Sinervo-Nature-380-240, 2002-Kerr-N-418-171, 2004-Kirkup-Nature-428-412,  2007-Reichenbach-N-488-1046, 2007-Reichenbach-PRL-99-238105}. This model has been shown to successfully describe some of the crucial features of simple  biological populations with interspecific competition, including certain populations of {\it E. coli} bacteria \cite{2002-Kerr-N-418-171, 2004-Kirkup-Nature-428-412} and lizards \cite{1996-Sinervo-Nature-380-240}.

Several extensions of and modifications to the RPS model incorporating further  species \cite{ 2008-Szabo-PRE-77-011906, 2012-Avelino-PRE-86-036112, 2012-Avelino-PRE-86-031119, 2013-Vukov-PRE-88-022123, 2014-Avelino-PLA-378-393, 2014-Avelino-PRE-89-042710, 2017-Avelino-PLA-381-1014, 2017-Brown-PRE-96-012147, 2019-Park-Chaos-29-051105, 2020-Avelino-PRE-101-062312}, interactions \cite{2004-Szabo-JPAMG-31-2599, 2009-Zhang-PRE-79-062901, 2010-Yang-C-20-2, 2014-Cianci-PA-410-66, 2014-Laird-Oikos-123-472, 2014-Rulquin-PRE-89-032133, 2015-Szolnoki-NJP-17-113033, 2016-Szolnoki-PRE-93-062307, 2017-Park-SR-7-7465,   2019-Park-EPL-126-38004, 2021-Bazeia-CSF-151-111255}, as well as biases which can favor some species over the others \cite{2001-Frean-PRSLB-268-1323, 2009-Berr-PRL-102-048102, 2019-Avelino-PRE-100-042209, 2019-Menezes-EPL-126-18003, 2020-Avelino-PRE-101-062312, 2020-Liao-N-11-6055, 2021-Avelino-EPL-134-48001}, have been investigated in the literature. In particular, the switch from unidirectional predator-prey interactions --- characteristic of the standard RPS model --- to bidirectional ones has been shown to have an enormous impact on the population dynamics, affecting both the evolution of the characteristic size of the spatial domains associated to the different species as well as the properties of the corresponding spatial patterns \cite{2012-Avelino-PRE-86-031119,2012-Avelino-PRE-86-036112}. 

In the RPS model, and most of its generalizations, the role of an individual of a given species as predator or prey only depends on the species of the individual with which it interacts, and that dependence is fixed {\it a priori}. However, in real systems the predator-prey role can also be affected by the relative strength of interacting individuals (e.g. adult prey may attack juvenile predators \cite{ctx7649326810003711,2012NatSR...2E.728C}) or by environmental factors which may trigger changes in population densities \cite{1988Sci...242...62B,2014RSOS....140186S}. 

In this letter we consider a generalization of the standard spatial stochastic RPS model --- which we shall refer to as the $\kappa$RPS model --- allowing for predator-prey reversal. Its main feature is the  incorporation of bidirectional predator-prey interactions between all the species, in addition to the unidirectional predator-prey interactions of the standard RPS model. Although in real systems the distinctive profile of interacting individuals --- beyond that characteristic of their own species --- is expected to affect the probability of occurrence of predator-prey reversal, in the present paper we shall treat such probability as uniform and investigate the corresponding impact on the population dynamics. Although this simplifying assumption can have some impact locally, it is not expected to strongly affect the overall dynamics assuming that the spatial distribution of the individuals of each species is not particularly sensitive to their distinctive profile.

\section{$\kappa$RPS Model}

Here, we shall describe the May-Leonard formulation of the spatial stochastic $\kappa$RPS model investigated in this letter. Consider a square lattice with $N^2$ sites with periodic boundary conditions in which individuals of all three species $i=1,2,3$ are initially randomly distributed. In a four-state May-Leonard formulation each site is either occupied by a single individual or left empty. The density of individuals of the species $i$ shall be denoted by $\rho_i = I_i/N^2$ while the density of empty sites will be referred to as $\rho_0 = I_0/N^2$ ($I_i$ and $I_0$ represent, respectively, the total number of individuals of the species $i$ and the total number of empty sites).

\begin{figure}[!t]
	\centering
	\includegraphics[scale=1.0]{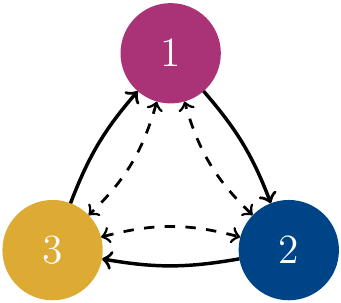}
	\caption{Predator-prey interactions scheme of the $3$ species predator-prey $\kappa$RPS model investigated in the present paper. The solid lines represent unidirectional predator-prey interactions characteristic of the standard RPS model while the dashed lines illustrate additional bidirectional predator-prey interactions featured in the $\kappa$RPS model.}
	\label{fig1}
\end{figure}

At each time step, a site occupied by an individual of one of the species $i$ (active) and one of its adjacent neighboring sites (passive) are randomly selected --- here, we shall consider a von Neumann neighborhood, composed of the active central cell and its four adjacent sites. One of three actions --- mobility, reproduction, or predation is then randomly selected --- with probabilities $m$, $r$, and $p$, respectively.

Whenever mobility is selected the active individual moves to the passive site and the individual at the passive site (if it exists) moves to the active one. That is:
\begin{equation}
i\ \odot \to \odot \ i\,, \nonumber
\end{equation}
where $\odot$ represents either an individual of any of the three different species or an empty site. If reproduction is selected and the passive site is empty, a new individual of the same species of the active individual is generated at that site:
\begin{equation}
i\ 0 \to i\ i\,. \nonumber
\end{equation}
Lastly, if predation is selected and the passive site is occupied by a prey of the individual of the species $i$ at the active site, a predator-prey  interaction is performed and the passive site is left empty:
\begin{equation}
i\ j \to i\ 0\,. \nonumber
\end{equation}
An individual of the species $j$ at the passive site is considered a prey of the species $i$ if $j=i+1$ (with probability $1-\kappa$) or if $j \neq i$ (with probability $\kappa$) --- see Fig. \ref{fig1} for an illustration of the two possibilities considered in the present paper. Here, $\kappa \in [0,1]$ is the single extra parameter of the $\kappa$RPS model with respect to the standard RPS model --- the standard RPS model is recovered if $\kappa=0$. Notice that, predator-prey reversal will occur with probability $\kappa$, whenever the passive site is occupied by an individual of the species $j-1$. Here, we are assuming modular arithmetic --- the integers $i$ and $j$ represent the same species whenever $i = j \mod 3$, where mod denotes the modulo operation.

If an interaction cannot be executed, which may happen when a reproduction interaction is selected and the passive is not an empty site or when a predator-prey interaction is selected and the passive is not occupied by a prey of the active individual, the procedure described in the last two paragraphs is repeated until a possible interaction is performed and the time step completed. A generation time (our time unit) is defined as the time necessary for $N^2$ successive mobility, 
reproduction or predator-prey interactions to be completed.







\begin{figure}[!t]
	\centering
	\includegraphics[scale=1.0]{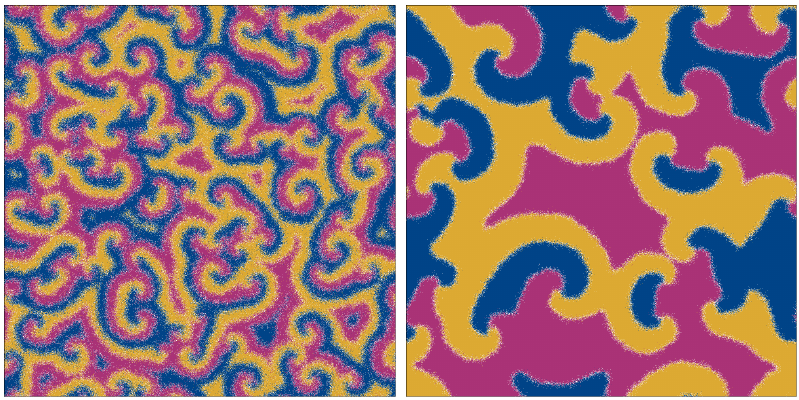}
	\caption{Snapshots of the spatial distribution of the different species at $t=10^4$ for a May-Leonard implementation of the spatial stochastic $\kappa$RPS model with $\kappa=0$ (left panel) and $\kappa=0.65$ (right panel) and random initial conditions. Notice that the size of the spiral patterns is significantly larger for $\kappa=0.65$ than for $\kappa=0$.}
	\label{fig2}
\end{figure}

\begin{figure}[!t]
	\centering
	\includegraphics[scale=1.5]{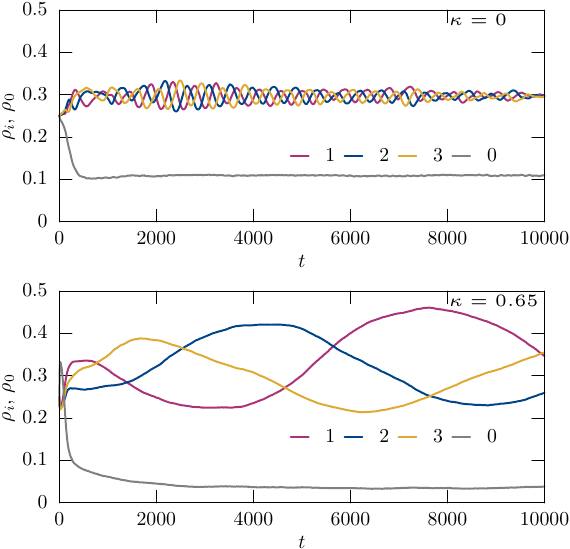}
	\caption{Time evolution of the densities $\rho_i$ and $\rho_0$ over time for the realizations of the spatial stochastic $\kappa$RPS model considered in Fig. \ref{fig2}. Notice that the characteristic oscillation time and fluctuations of the species densities are significantly larger for $\kappa=0.65$ than for $\kappa=0$. Also, the relaxation of $\rho_0$ towards a nearly constant value is faster for $\kappa=0$.}
	\label{fig3}
\end{figure}


\section{Population dynamics}

In this section we shall describe the results of spatial stochastic lattice based numerical simulations of the $\kappa$RPS model with random initial conditions --- at the start of the simulations each site was filled with an individual of any of the 3 species or left empty with a uniform discrete probability of $1/4$. The following parameters were used in all simulations: $m=0.8$, $p=0.1$, $r=0.1$. Also, $N=1000$ except in the case of the simulations used to produce Figs. \ref{fig6},  \ref{fig7} and \ref{fig8} (for which $N=250$).

Figure \ref{fig2} presents snapshots of the spatial distribution of the individuals of different species taken after $10^4$ generations for $\kappa=0$ (left panel) and $\kappa=0.65$ (right panel). Although a network of spiral patterns emerges in both cases, their characteristic length is significantly larger for $\kappa=0.65$ than for $\kappa=0$ (the latter corresponds to the standard spatial stochastic RPS model),  thus showing that for sufficiently large values of $\kappa$ predator-prey reversal can have a significant impact on the overall dynamics.

\begin{figure}[!t]
	\centering
	\includegraphics[scale=1.05]{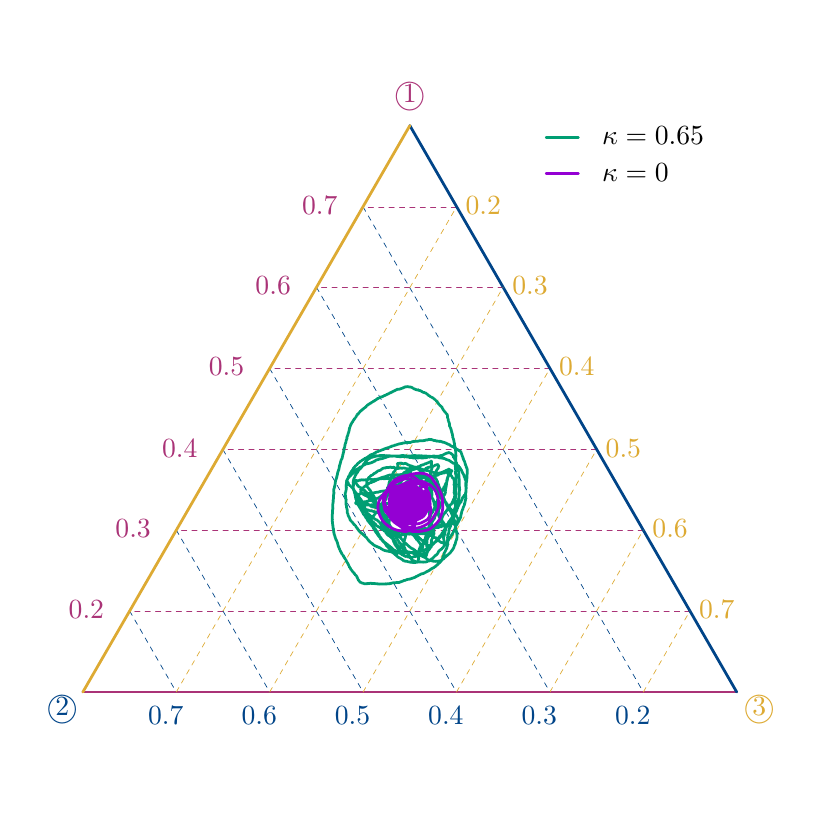}
	\caption{Phase space evolution for the realizations of the spatial stochastic $\kappa$RPS model considered in Fig. \ref{fig2}. Notice that the phase space area explored in a fixed interval of time is significantly larger for $\kappa=0.65$ than for $\kappa=0$.}
	\label{fig4}
\end{figure}

\begin{figure}[!t]
	\centering
	\includegraphics[scale=1.0]{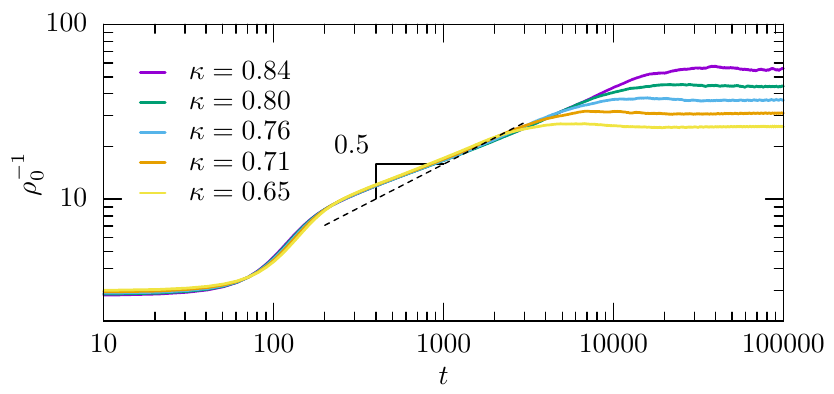}
	\caption{Time evolution of $\rho_0 ^{-1}$ (proportional to the characteristic length of the network $L$) for spatial stochastic $\kappa$RPS models with $\kappa=0.65$, $0.71$, $0.76$, $0.80$ and $0.84$ --- each point on the two curves was estimated from an average of $10^2$ simulations with random initial conditions. Notice the increase with $\kappa$ of the asymptotic value of $\rho_0^{-1}$ and of the transition time between the transient and asymptotic scaling regimes.}
	\label{fig5}
\end{figure}

This impact is quantified in Figure \ref{fig3} where the evolution of the densities $\rho_i$ and $\rho_0$ of the different species and empty sites with time is presented for the realizations of the spatial stochastic $\kappa$RPS model considered in Fig. \ref{fig2}. It shows, in particular, that the larger characteristic size of the structures observed for $\kappa=0.65$ is associated with  a larger characteristic oscillation time and larger fluctuations of the species densities (compared to the $\kappa=0$ case). Also, the relaxation of $\rho_0$ towards a nearly constant value is slower for $\kappa=0.65$ than for $\kappa=0$. Figure \ref{fig4} displays the same phase space evolution on a triangular plot, but considering a significantly larger time span ($10^5$ generations). It again shows that the region of phase space explored in a fixed interval of time is significantly larger for  $\kappa=0.65$ (compared to the $\kappa=0$ case). 

Predation interactions occur mostly at the borders between different domains. As a result, the empty sites are mainly concentrated at such borders, defining an interface network whose average thickness is time independent and whose total length $L_{\rm T}$ is roughly proportional to the density of empty spaces $\rho_0$. The characteristic length of the network may then be defined as $L \equiv A/L_{\rm T}$, where $A$ is the total of the simulation box, and satisfies $L \propto \rho_0 ^{-1}$ \cite{2012-Avelino-PRE-86-031119}.


Figure \ref{fig5} shows the time evolution of $\rho_0 ^{-1}$ for spatial stochastic $\kappa$RPS models with $\kappa=0.65$, $0.71$, $0.76$, $0.80$ and $0.84$. Each point on the curves was estimated from the average of $10^2$ simulations with random initial conditions. In all cases there is a relatively long transient scaling regime given approximately by 
\begin{equation}
L \propto \rho_0^{-1} \propto t^{1/2}\,, \label{scaling1}
\end{equation}
before an asymptotic scaling regime with 
\begin{equation}
L = L_{\rm [assymptotic]} \propto \rho_0^{-1} = \rho^{-1}_{\rm 0 [assymptotic]} =\rm const\,. \label{scaling2}
\end{equation}
Notice that both the asymptotic characteristic length and the duration of the transient scaling regime are  increasing functions of $\kappa$. In fact they are directly related, since the time of the transition $t_{\rm transition}$ between the transient and asymptotic scaling regime scales roughly as 
\begin{equation}
t_{\rm transition} \propto \rho^{-2}_{\rm 0[assymptotic]} \propto L_{\rm [assymptotic]}^2\,. \label{scaling3}
\end{equation}
We shall provide provide a simple model for the dependence of the asymptotic characteristic length and of the duration of the transient scaling regime on $\kappa$ in the next section. 

\begin{figure}[!t]
	\centering
	\includegraphics[scale=1.0]{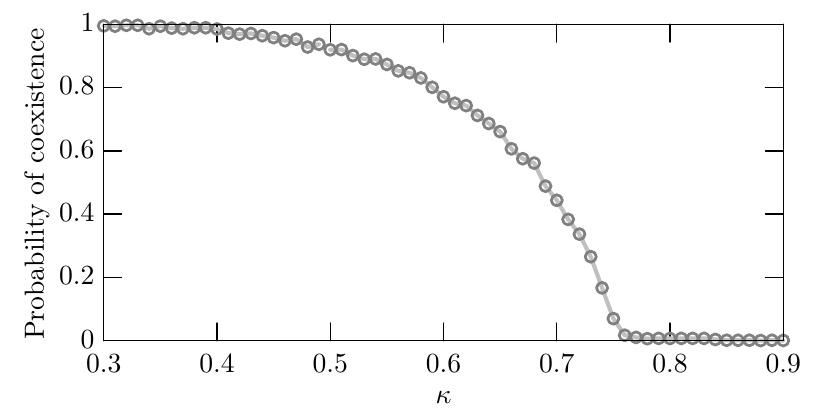}
	\caption{Probability of coexistence as a function of the value of the parameter $\kappa$. Each point was estimated from $2 \times 10^3$ simulations with a total simulation time equal to $10^5$ generations. Notice that coexistence is expected to be maintained for $\kappa < 0.4$ while for $\kappa > 0.7$ extinction of two of the three species is the most likely outcome.}
	\label{fig6}
\end{figure}


If the simulation lattices are not sufficiently large then predator-prey reversal can have a significant negative impact on coexistence. This will happen whenever the expected asymptotic characteristic length is of the same order or larger than the box size. In order to investigate this we performed simulations of the $\kappa$RPS models with $\kappa=0$ and $\kappa=0.65$ considered in Fig. \ref{fig6}, but now using smaller simulation boxes (with $N=250$ instead of $N=1000$). 

Figure \ref{fig6} displays the probability that the three species still coexist after $10^5$ generations estimated from these simulations as a function of the value of the parameter $\kappa$ --- each point was estimated from $2 \times 10^3$ simulations. It shows that for the considered model parameters coexistence is expected to be maintained for $\kappa < 0.4$, while for $\kappa > 0.7$ extinction of two of the three species in less than $10^5$ generations is the most likely outcome.

\section{Dynamics of circular domains}

Here we shall consider the dynamics of a population of a single predator species 1, initially confined to a circular spatial domain of radius $R$ surrounded by a territory occupied by a single species 2 (the prey of species 1). This simple model incorporates two opposite contributions: i) standard predator-prey interactions which contribute to increase to the expansion of the predator population of the inner domain; ii) the possibility of predator-prey reversal which has the opposite effect. 

\begin{figure}[!t]
	\centering
	\includegraphics[scale=1.0]{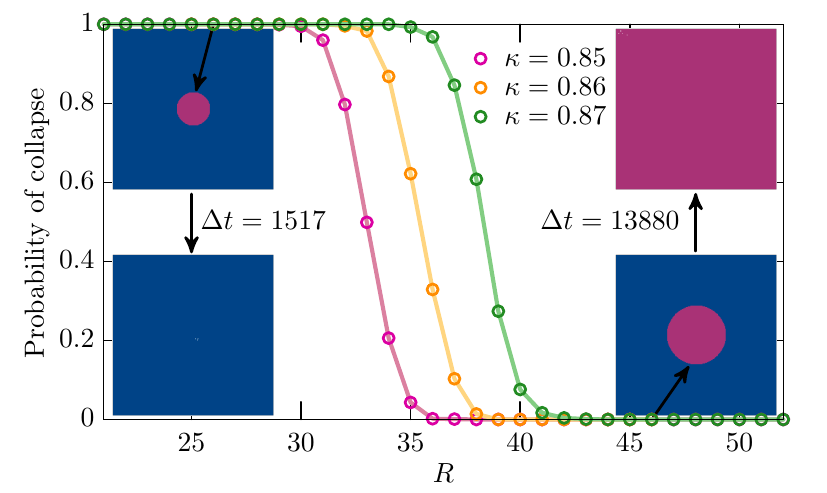}
	\caption{The probability $P$ of collapse of the region dominated by the predator species 1 (initially confined to a circular spatial domain of radius $R$) as a function of $R$. Each point was estimated from $10^3$  simulations. The left and right inset panels show two snapshots of runs with initial radius $R = 26$ and $R = 46$, respectively. The times required for the initially circular domain to collapse (left inset panels) or to invade all the territory (right inset panels) are displayed between snapshots (lower and upper inset panels represent the initial and final configurations).}
	\label{fig7}
\end{figure}

Figure \ref{fig7} shows the probability $P$ of collapse of the region dominated by the predator species 1 as a function of $R$ for three different values of $\kappa$. Each point was estimated from $10^3$ simulations. The left and right inset panels show two snapshots of runs with $\kappa=0.86$ and initial radius $R = 26$ and $R = 46$, respectively. The times required for the initially circular domain to collapse (left inset panels) or to invade all the territory (right inset panels) are displayed between snapshots (lower and upper inset panels represent the initial and final configurations). Figure \ref{fig7} shows that the larger the dynamical relevance of predator-prey reversal (which increases with $\kappa$), the larger the value of $R$ for which the extinction probability of the inner and outer species coincide (corresponding to collapse probability equal to 0.5). 

For $\kappa=1$ the dynamics of the $\kappa$RPS model is  curvature driven, with the velocity of the interfaces separating different domains being roughly proportional to the inverse of their curvature radii ($v \propto R^{-1}$) \cite{2012-Avelino-PRE-86-031119} --- the proportionality constant being dependent on the values of $r$, $p$ and $m$. On the other hand, for $\kappa=0$ one recovers the standard spatial stochastic RPS dynamics in which the interfaces separating different domains have a roughly constant velocity (but dependent on the values of $r$, $p$ and $m$) irrespectively of their curvature radii. For $0 < \kappa < 1$ the $\kappa$RPS model essentially combines the two effects with weight $\kappa$ ($\kappa=1$ model) and $1-\kappa$ ($\kappa=0$ model). Hence, one would expect $R_*$, the value of $R$ for which $P=0.5$, to scale as
\begin{equation}
R_* \propto \frac{\kappa}{1-\kappa}\,. \label{rstar}
\end{equation}

\begin{figure}[!t]
	\centering
	\includegraphics[scale=1.0]{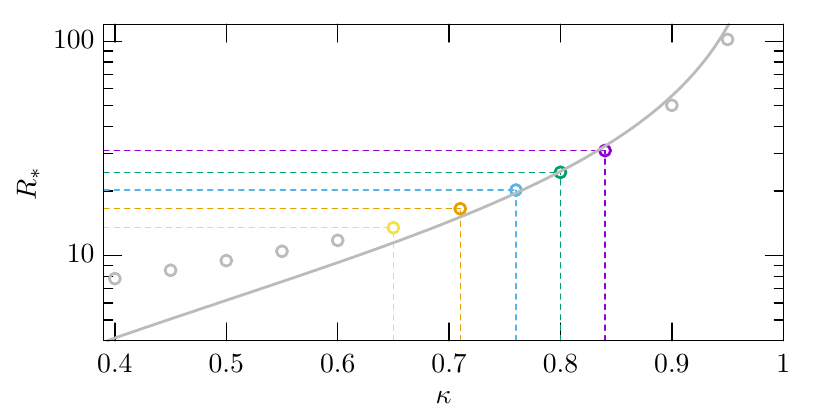}
	\caption{The value of $R_* \equiv R(P=0.5)$ as a function of $\kappa$. The open circles and the grey solid line represent respectively the  numerical results and the fit obtained using Eq.~(\ref{rstar}). Notice the good agreement between the analytical and numerical results for $\kappa \gsim 0.7$. The open circles corresponding to the values of $\kappa$ considered in Fig. \ref{fig5} are crossed by dashed lines for an easier comparison with the results presented therein. 
	}
	\label{fig8}
\end{figure}

Figure \ref{fig8} displays the value of $R_* \equiv R(P=0.5)$ as a function of $\kappa$. The numerical results (open circles) and the fit obtained using Eq.~(\ref{rstar}) (black solid line) are in good agreement for $\kappa \gsim 0.7$. This is no longer true for smaller values of $\kappa$ (corresponding to $R_* \lsim 10$), a regime where finite spatial-grid effects are expected to significantly affect the results. 

The dashed lines passing through the open circles corresponding to the values of $\kappa$ considered in Fig. \ref{fig5} allow for easier comparison with the results presented therein. Notice that the ratios between the values of the asymptotic characteristic lengths $L \propto \rho_0^{-1} = \rm const$ obtained for the different $\kappa$ in Fig. \ref{fig5} roughly coincide with the expectation based on the ratios of the corresponding values of $R_*$. 

\section{Conclusions}

In this letter we investigated the dynamics of the $\kappa$ RPS model, a single parameter generalization of the standard three species RPS model allowing for predator-prey reversal. We have shown that for $\kappa <1$ and sufficiently large simulation boxes, a population network characterized by a spiral patterns qualitatively similar to that of the standard RPS model eventually emerges. However, we have found that the emergence of such network is in general preceded by a transient curvature dominated scaling regime in which the characteristic length of the population network grows with time (roughly proportionally
to $t^{1/2}$). We have provided a simple model, based on the dynamics of circular domains, to quantify the dependence of the asymptotic value of the characteristic length on $\kappa$ and shown that if the simulation boxes are not sufficiently large then predator-prey reversal can have a significant negative impact on coexistence.

\acknowledgments
P.P.A. acknowledges the support by Fundação para a Ciência e a Tecnologia (FCT) through the research grants UIDB/04434/2020, UIDP/04434/2020. B.F.O. and R.S.T. thanks CAPES-Finance Code 001, Funda\c c\~ao Arauc\'aria, and INCT-FCx (CNPq/FAPESP) for financial and computational support.


\end{document}